\documentclass[aps]{revtex4}
\usepackage{amsmath}
\usepackage{amssymb}

\begin{document}

\title{Spin-charge separation for the SU(3) gauge theory}
\author{Vladimir Dzhunushaliev
\footnote{Senior Associate of the Abdus Salam ICTP}}
\email{dzhun@krsu.edu.kg}
\affiliation{Dept. Phys. and Microel. Engineer., Kyrgyz-Russian Slavic University, 
Bishkek, Kievskaya Str. 44, 720021, Kyrgyz Republic \\
and \\
Institut f\"ur Physik, Humboldt--Universit\"at zu Berlin,
Newtonstr. 15, D-12489  Berlin, Germany
}


\begin{abstract}
The idea of a spin-charge separation of the SU(2) gauge potential is extended to the SU(3) case. It
is shown that in this case there exist different non-perturbative ground states characterized by
different gauge condensate $A^B_\mu A^B_\mu \neq 0$. 
\end{abstract}

\maketitle

\section{Introduction}

One of the main problems in quantum field theory is the quantization of strongly interacting felds. 
In quantum chromodynamics this problem leads to the fact that up to now we do not
completely understand the confinement of quarks. Mathematically the problem is connected with
quartic term 
$g^2 \left ( f^{ABC} A^B_\mu A^C_\nu \right )^2$ in the SU(3) Lagrangian: we have no exact 
mathematical
tools for 
the non-perturbative path integration of such non-quadratic Lagrangian. 
\par 
In this case it is useful to have any analogy with other area of physics. In Ref. \cite{Niemi:2005qs} the 
authors considered the similarity between High-T$_c$ cuprate superconductivity in condensed matter
physics 
and the problem of a mass gap in the Yang-Mills theory. The authors suggest that in both cases the
basic 
theoretical problems is the absense of a natural condensate to describe the symmetry breaking. 
The method which is applied in this investigation is a slave-boson decomposition \cite{Anderson:1987gf} - 
\cite{lee}.  
\par 
In Ref. \cite{Chernodub:2005jh} the idea is presented that an analogy may exist between the SU(2) 
Yang-Mills theory in the low-temperature phase and a nematic liquid crystal. The idea is based on a
spin-charge 
separation of the gluon field in the Landau gauge.
\par 
In Ref. \cite{Dzhunushaliev:2000ma} the idea is proposed that in  High-T$_c$ superconductivity may
exist an
analog of a hypothesized flux tube between quarks in quantum chromodynaimcs where such flux tube
essentially increases the interaction energy of two interacting quarks in comparison with the
interaction energy for two electrons.
\par 
In this paper we would like to investigate such spin-charge separation for the SU(3) gauge field theory 
and additionally to show that the SU(2) gauge field theory may have another spin-charge separation.

\section{Spin-charge separation}

In the matrix theory \cite{roger} there exists the theorem that any real $(m \times n), m > n$
matrix $A$ can be decomposed as 
\begin{equation}
	A = QR
\label{1-10}
\end{equation}
where $Q$ is an $(m \times n)$ orthogonal matrix ($Q^T Q = 1$) and $R$ is $(n \times n)$ upper
triangular matrix. 
If $A$ is $(m \times n), m < n$ then $Q$ is an $(m \times m)$ orthogonal matrix and $R$ is 
$(m \times n)$ upper triangular matrix. Following to this theorem, we can decompose any SU(2) gauge
component $A^a_\mu$ as 
\begin{equation}
	A_\mu^{\text{   } a} = \tilde{e}_\mu^{\text{   } i} \tilde \Phi^{ia}
\label{1-20}
\end{equation}
where $a=1,2,3$ is the SU(2) color index and enumerates the columns; $\mu =1,2,3,4$ (we consider the
Euclidean version of the theory) and enumerates the rows; $i=1,2,3$ is an inner index which
enumerates the columns. Let us introduce the unity 
\begin{equation}
	1 = \Lambda \Lambda^{-1}
\label{1-30}
\end{equation}
where $\Lambda$ is an SO(3) orthogonal matrix. The unity can be inserted in Eq. \eqref{1-20} by
such a way that 
\begin{equation}
	A_\mu^{\text{   } a} = \left ( 
		\tilde{e}_\mu^{\text{    } i} \Lambda^{ij}
	\right )
	\left (
		\Lambda^{kj} \tilde \Phi^{ka}
	\right ) = 
	e_\mu^{\text{    } i} \Phi^{ia} 
\label{1-40}
\end{equation}
where $e_\mu^{\text{    } i} = \tilde{e}_\mu^{\text{    } j} \Lambda^{ji}$, 
$\Phi^{ia} = \Lambda^{ji} \tilde \Phi^{ja}$. This decomposition is the subject of the investigation
in Ref. \cite{Chernodub:2005jh}. The matrix
$A_\mu^{\text{   } a}$ is a $(4 \times 3)$ matrix, $\tilde{e}_\mu^{\text{   } i}$ is a 
$(4 \times 3)$ matrix and $\tilde \Phi^{ka}$ is a $(3 \times 3)$ matrix. 
\par 
In Ref. \cite{Chernodub:2005jh} the idea is presented that the SU(2) Yang - Mills theory can be
associated with a nematic crystal in which the "molecules" are directed in the internal SO(3)
space. The adjoint "matter" field 
\begin{equation}
	\chi^{ij} = \sum \limits_a \Phi^{ai} \Phi^{aj}
\label{1-50}
\end{equation}
can be associated with the dielectric susceptibility 
\begin{equation}
	\tilde \chi_{\alpha \beta} = \Delta \tilde \chi \sum \limits_s n^{(s)}_\alpha n^{(s)}_\beta 
\label{1-60}
\end{equation}
where $n^{(s)}_\alpha$ is the direction of the axis of the s$^{th}$ molecule; 
$\Delta \tilde \chi = \tilde \chi_\parallel - \tilde \chi_\perp$ is the anisotropy in the
diamagnetic susceptibility along and perpendicular to the molecule axis. 

\section{Another decomposition of SU(2) gauge fields}
\label{another}

One can present the potential $A^a_{\text{   }\mu}$ also as a $(3 \times 4)$ matrix where $a$
enumerates
the rows and $\mu$ -- the columns. Then the corresponding decomposition will be 
\begin{equation}
	A^a_{\text{   }\mu} = \tilde \Phi^{ai} \tilde e^i_{\text{   }\mu}
\label{2-10}
\end{equation}
where $\tilde \Phi^{ai}$ is the orthogonal matrix $\tilde \Phi^{ai} \tilde \Phi^{aj} = \delta^{ij}$
and $\tilde e^i_{\text{   }\mu}$ is an upper triangular matrix. Again we can insert the unity 
$1 = \Lambda \Lambda^{-1}$ between $\tilde \Phi$ and $\tilde e$ on the r.h.s. of Eq. \eqref{2-10}.
Finally we have 
\begin{equation}
	A^a_{\text{   }\mu} = \Phi^{ai} e^i_{\text{   }\mu}
\label{2-20}
\end{equation}
where $\Phi = \tilde \Phi \Lambda$ and $e = \Lambda^{-1} \tilde e$. Now we would like to rewrite
the SU(2) Lagrangian in terms of the fields $\Phi^{ai}$ and $e^i_{\text{   } \mu}$ similar to Ref.
\cite{Chernodub:2005jh}. The field strength $F^a_{\text{  } \mu \nu}$ is 
\begin{equation}
\begin{split}
	F^a_{\text{  } \mu \nu} = & \partial_\mu A^a_{\text{   } \nu } - 
	\partial_\nu A^a_{\text{   } \mu } + 
	g \epsilon^{abc} A^b_{\text{   } \mu } A^c_{\text{   } \nu } = 
	\\
	&
	\Phi^{ai} \left (
		\partial_\mu e^i_{\text{   } \nu } - \partial_\nu e^i_{\text{   } \mu } 
	\right ) + 
	\left (
		e^i_{\text{   } \nu } \partial_\mu \Phi^{ai} - e^i_{\text{   } \mu } \partial_\nu \Phi^{ai} 
	\right ) + 
	g \epsilon^{abc} \Phi^{bi} \Phi^{cj} e^i_{\text{   } \mu } e^j_{\text{   } \nu }
\label{2-30}
\end{split}
\end{equation}
where $\epsilon^{abc}$ are the SU(2) structural constants. The terms without coupling constant $g$
can be rewritten as 
\begin{equation}
\begin{split}
	&\Phi^{ai} \left (
		\partial_\mu e^i_{\text{   } \nu } - \partial_\nu e^i_{\text{   } \mu } 
	\right ) + 
	\left (
		e^i_{\text{   } \nu } \partial_\mu \Phi^{ai} - e^i_{\text{   } \mu } \partial_\nu \Phi^{ai} 
	\right ) = 
	\\ 
	&
	\Phi^{bi} \left [
		\delta^{ab} \partial_\mu e^i_{\text{  } \nu} + 
		\frac{1}{2} \left (
			e^i_{\text{  } \nu} \partial_\mu \Phi^{aj} - 
			e^j_{\text{  } \mu} \partial_\nu \Phi^{ai} 
		\right ) \Phi^{bj}
	\right ] - 
	\Phi^{bi} \left [
		\delta^{ab} \partial_\nu e^i_{\text{  } \mu} + 
		\frac{1}{2} \left (
			e^i_{\text{  } \mu} \partial_\nu \Phi^{aj} - 
			e^j_{\text{  } \nu} \partial_\mu \Phi^{ai} 
		\right ) \Phi^{bj}
	\right ] = 
	\\
	&
	\Phi^{bi} \left [
		D^{ab}_{\;\;\; \mu \nu} (\Gamma ) e^i_{\text{  } \nu} - 
		D^{ab}_{\;\;\; \nu \mu} (\Gamma ) e^i_{\text{  } \mu}
	\right ]
\label{2-40}
\end{split}
\end{equation}
where $D^{ab}_{\;\;\; \mu \nu} (\Gamma )$ is an analog of the covariant derivative with the
``connection'' $\Gamma$ 
\begin{equation}
	\Gamma^{ab,ij}_{\quad \mu \nu}\left ( e^i_{\nu} \right ) = 
	\frac{1}{2} \left (
		e^i_{\text{  }\nu} \partial_\mu \Phi^{aj} - e^i_{\text{  }\mu} \partial_\nu \Phi^{ai}
	\right )
\label{2-50}
\end{equation}
Then the SU(2) Lagrangian can be written as 
\begin{equation}
	L = L_0 + L_1 + L_2
\label{2-60}
\end{equation}
with 
\begin{eqnarray}
	L_0 &=& \frac{1}{4} \left \{
		\Phi^{bi} \left [
			D^{ab}_{\;\;\; \mu \nu} (\Gamma) e^i_{\text{  } \nu} - D^{ab}_{\;\;\; \nu \mu} (\Gamma)
e^i_{\text{ } \mu}
		\right ]
	\right \} ,
\label{2-70}\\
	L_1 &=& \frac{g}{2} \Phi^{bi} \left [
			D^{ab}_{\;\;\; \mu \nu} (\Gamma) e^i_{\text{  } \nu} - 
			D^{ab}_{\;\;\; \nu \mu} (\Gamma) e^i_{\text{  }\mu}
		\right ] 
		\epsilon^{abc} \Phi^{bk} \Phi^{cl} e^k_{\text{  } \mu} e^l_{\text{  }\nu} ,
\label{2-80}\\
	L_2 &=& \frac{g^2}{4} \left [
		\left ( \mathrm{Tr} \; \chi \right )^2 - \; \mathrm{Tr} \left ( \chi \right )^2
	\right ]
\label{2-90}
\end{eqnarray}
where 
\begin{equation}
	\chi^{ij} = e^i_{\text{  } \mu} e^j_{\text{  } \mu} .
\label{2-100}
\end{equation}
Similar to Ref. \cite{Chernodub:2005jh} the quantity $\chi^{ij}$ can be associated with the
nematic crystal with one difference: the "molecules" are directed in the Euclidean space-time with
the coordinates $x^\mu$. Absolutely by the same way as in Ref. \cite{Chernodub:2005jh} one can
calculate the ground state of the nematic associated with the Yang - Mills theory
\eqref{2-70}-\eqref{2-90}. If we introduce the eigenvalues of the matrix 
$\chi = \mathrm{diag} \left \{ \chi_1, \chi_2, \chi_3 \right \}$, the ground state 
$\chi = \chi_0$ is defined as 
\begin{equation}
	\sum \limits_{i,j=1}^4 \chi^{(0)}_i \chi^{(0)}_j = 0
\label{2-110}
\end{equation}
with the constraints 
\begin{equation}
	\sum \limits_{i=1}^4 \chi^{(0)}_i \ge 0, \quad 
	\prod \limits_{i=1}^4 \chi^{(0)}_i \ge 0. 
\label{2-120}
\end{equation}
The solutions of \eqref{2-110} \eqref{2-120} are given as 
\begin{equation}
	\chi^{(0)}_1 = \chi^{(0)}_2 = 0, \quad \chi^{(0)}_3 \ge 0
\label{2-130}
\end{equation}
The most interesting in this consideration is a non-perturbative vacuum which corresponds to 
$\chi^{(0)}_3 \neq 0$. Clearly, this vacuu, state is a $A^2$--condensate 
\begin{equation}
	\left \langle A^a_\mu A^a_\mu \right \rangle = \chi^{(0)}_3 \neq 0 .
\label{2-140}
\end{equation}

\section{SU(3) spin-charge separation}

In this section we would like to repeat the SU(2) matrix decomposition of the previous section for
the SU(3) case. 

\subsection{$A^{\text{  }B}_\mu$ gauge potential as a $(4 \times 8)$ matrix}

This case is similar to the spin-charge separation used in Ref. \cite{Chernodub:2005jh} 
\begin{equation}
	A^{\text{  }B}_\mu = e^{\text{  }i}_\mu \Phi^{iB}
\label{3-1-10}
\end{equation}
The matrix $e^{\text{  }i}_\mu$ is orthogonal one 
$e^{\text{  }i}_\mu e^{\text{  }j}_\mu = \delta^{ij}; i,j = 1,2,3,4$. The matrix
$e^{\text{  }i}_\mu$ is similar to the 4-bein but with one essential difference. Generally
speaking, one has 
\begin{equation}
	e^{\text{  }i}_\mu e^{\text{  }i}_\nu \neq \delta_{\mu \nu}
\label{3-1-20}
\end{equation}
Using this decomposition, one can write 
\begin{equation}
	L_{SU(3)} = \frac{1}{4} \left (F^a_{\mu \nu} \right )^2 = L_0 + L_1 + L_2
\label{3-1-30}
\end{equation}
with 
\begin{eqnarray}
	L_0 &=& \frac{1}{2} \left ( D_\mu \phi^{iB} \right )^2 + 
	\frac{1}{8} \phi^{lB} \Phi^{jB} \left (
		\partial_\mu e^{\text{  }j}_\nu - \partial_\nu e^{\text{  }j}_\mu
	\right )
	\left [
		\left ( \partial_\mu e^{\text{  }l}_\nu - \partial_\nu e^{\text{  }l}_\mu \right ) - 
		e^{\text{  }i}_\nu e^{\text{  }i}_\alpha 
		\left ( \partial_\mu e^{\text{  }l}_\alpha - \partial_\alpha e^{\text{  }l}_\mu \right ) 
	\right ] - 
\nonumber \\
	&&\frac{1}{2} \left [
		e^{\text{  }i}_\nu \partial_\mu \Phi^{iB} + \frac{1}{2} \left (
				\partial_\mu e^{\text{  }i}_\nu - \partial_\nu e^{\text{  }i}_\mu 
			\right )
	\right ] 
	\left [
		e^{\text{  }j}_\mu \partial_\nu \Phi^{jB} + \frac{1}{2} \left (
				\partial_\nu e^{\text{  }j}_\mu - \partial_\mu e^{\text{  }j}_\nu 
			\right )
	\right ] ,
\label{3-1-40}\\
	L_1 &=& \frac{g}{2} f^{BCD} \left \{
		\left [
			e^{\text{  }i}_\nu  \partial_\mu \Phi^{iB} + 
			\frac{1}{2} \left (
				\partial_\mu e^{\text{  }i}_\nu - \partial_\nu e^{\text{  }i}_\mu
				\right ) \Phi^{iB}
		\right ] - 
		\biggl [ \mu \leftrightarrow \nu \biggl ]
	\right \}
	\Phi^{j C} \Phi^{k D} e^{\text{  }j}_\mu e^{\text{  }k}_\nu ,
\label{3-1-50}\\
	L_2 &=& \frac{g^2}{4} f^{BCD} f^{BMN} A^C_\mu A^D_\nu A^M_\mu A^N_\nu = 
	\frac{g^2}{4} \chi^{CM} f^{BCD} f^{BMN} \chi^{DN} = 
	- \frac{g^2}{4} \mathrm{Tr} \left (
		\Phi f^B \Phi^T
	\right )^2
\label{3-1-60}
\end{eqnarray}
where $f^B$ is the matrix $\left ( f^B \right )^{MN}$ and 
\begin{equation}
	\chi^{AB} = \Phi^{iA} \Phi^{iB} .
\label{3-1-70}
\end{equation}
The covariant derivative $D_\mu \phi^{iB}$ is defined in the following way 
\begin{equation}
	D_\mu \phi^{iB} = \partial_\mu \Phi^{iB} + \Gamma(e)^{ij}_\mu \Phi^{jB}
\label{3-1-80}
\end{equation}
and the connection $\Gamma(e)$ as 
\begin{equation}
	\Gamma^{ij}_\mu (e) = e^{\;\; i}_\nu \left (
		\partial_\mu e^{\;\; j}_\nu - \partial_\nu e^{\;\; j}_\mu
	\right )
\label{3-1-90}
\end{equation}
In order to find possible vacuum state we should to find the values of the condensate 
$A^B_\mu A^B_\mu = \mathrm{Tr} \chi$ for which the potential term $L_2$ is zero. Let the matrix
$\chi^{AB}$ is diagonalized 
\begin{equation}
	\chi^{AB} = \mathrm{diag} \left \{
		\chi_1, \cdots, \chi_8
	\right \}.
\label{3-1-100}
\end{equation}
In this case 
\begin{equation}
\begin{split}
	L_2 = &\frac{g^2}{4} \left [
		2 \left ( \chi_1 \chi_2 +  \chi_1 \chi_3 + \chi_2 \chi_3 
	\right ) + 
	\right .
	\\ &
	\frac{1}{2} \left (
		\chi_1 \chi_4 + \chi_1 \chi_5 + \chi_1 \chi_6 + \chi_1 \chi_7 + 
	\chi_2 \chi_4 + \chi_2 \chi_5 + \chi_2 \chi_6 + \chi_2 \chi_7 + 
	\right .
	\\ &
	\chi_3 \chi_4 + \chi_3 \chi_5 + \chi_3 \chi_6 + \chi_3 \chi_7 +
		4 \chi_4 \chi_5 + \chi_4 \chi_6 + \chi_4 \chi_7 + 3 \chi_4 \chi_8 + 
	\\ &
	\left . 
	\left .
		\chi_5 \chi_6 + \chi_5 \chi_7 + 3 \chi_5 \chi_8 + 
		4 \chi_6 \chi_7 + \chi_6 \chi_8 + 
		3 \chi_7 \chi_8 
	\right ) 
	\right ]
\label{3-1-110}
\end{split}
\end{equation}
The first term in eq. \eqref{3-1-110} correspons to the SU(2) subgroup \eqref{2-90}. For the
perturbative vacuum the solution is 
\begin{equation}
	\chi_{i} = 0 , \quad i=1, \cdots , 8
\label{3-1-120}
\end{equation}
The possible non-perturbative vacuum is more complicated then in the SU(2) case. One can exist
different vacuum states. The first vacuum state is similar to the SU(2) case and it is defined by
the relation 
\begin{equation}
	\chi_{i} = 0 , \quad \chi_j \neq 0,
\label{3-1-130}
\end{equation}
$j$ is a fixed number. In this case the vacuum condensate is given in the following manner 
\begin{equation}
	\mathrm{Tr} \chi = \left \langle 
		A^B_\mu A^B_\mu
	\right \rangle = \chi_j 
\label{3-1-135}
\end{equation} 
From eq. \eqref{3-1-120} we see that not all $\chi_i$ are equivalent that means that the
corresponding vacuum states may be nonequivalent in the contrast with the SU(2) case. 
\par 
The second possibility is the case when 
\begin{equation}
	\chi_{i} = 0 
\label{3-1-140}
\end{equation}
but, for example, three $\chi_{6,7,8} \neq 0$. In this case we have the following relation between
$\chi_{6,7,8}$ 
\begin{equation}
	\frac{4}{3} \chi_6 \chi_7 + \chi_6 \chi_8 + \chi_7 \chi_8 = 0 
\label{3-1-150}
\end{equation}
but $\chi_{6,7,8}$ are independent degrees of freedom and they can be do not satisfy the relation
\eqref{3-1-150}. Thus in this case it will be a \textit{special} vacuum state and the vacuum
\textit{special} condensate is 
\begin{equation}
	\left \langle A^B_\mu A^B_\mu \right \rangle = \chi^{AA} = 
	\sum \chi^i = 
	\chi_6 + \chi_7 - \frac{4}{3} \frac{\chi_6 \chi_7}{\chi_6 + \chi_7}
\label{3-1-160}
\end{equation}
Other cases with four and more non-zero $\chi_i$ can be considered analogously.

\subsection{$A^{\text{  }B}_\mu$ gauge potential as a $(8 \times 4)$ matrix}

In this case 
\begin{equation}
	A^{B}_{\text{   }\mu} = \Phi^{Bi} e^i_{\text{   }\mu} 
\label{3-2-10}
\end{equation}
where $\Phi^{Bi} \Phi^{Bj} = \delta^{ij}$. The same calculations as in the section \ref{another}
gives us 
\begin{equation}
\begin{split}
	F^B_{\text{  } \mu \nu} = & \partial_\mu A^B_{\text{   } \nu } - 
	\partial_\nu A^B_{\text{   } \mu } + 
	g f^{BCD} A^C_{\text{   } \mu } A^D_{\text{   } \nu } = 
	\\
	&
	\Phi^{Bi} \left (
		\partial_\mu e^i_{\text{   } \nu } - \partial_\nu e^i_{\text{   } \mu } 
	\right ) + 
	\left (
		e^i_{\text{   } \nu } \partial_\mu \Phi^{Bi} - e^i_{\text{   } \mu } \partial_\nu \Phi^{Bi} 
	\right ) + 
	g f^{BCD} \Phi^{Ci} \Phi^{Dj} e^i_{\text{   } \mu } e^j_{\text{   } \nu }
\label{3-2-20}
\end{split}
\end{equation}
and the SU(3) Lagrangian 
\begin{equation}
	L_{SU(3)} = L_0 + L_1 + L_2
\label{3-2-30}
\end{equation}
can be written as 
\begin{eqnarray}
	L_0 &=& \frac{1}{4} \left \{
		\Phi^{Bi} \left [
			D^{AB}_{\;\;\; \mu \nu} (\Gamma) e^i_{\text{  } \nu} - D^{AB}_{\;\;\; \nu \mu} (\Gamma)
e^i_{\text{ } \mu}
		\right ]
	\right \} ,
\label{3-2-40}\\
	L_1 &=& \frac{g}{2} \Phi^{Bi} \left [
			D^{AB}_{\;\;\; \mu \nu} (\Gamma) e^i_{\text{  } \nu} - 
			D^{AB}_{\;\;\; \nu \mu} (\Gamma) e^i_{\text{  }\mu}
		\right ] 
		f^{ABC} \Phi^{Bk} \Phi^{Cl} e^k_{\text{  } \mu} e^l_{\text{  }\nu} ,
\label{3-2-50}\\
	L_2 &=& \frac{g^2}{4} \left ( f^{BCD} f^{BMN} \right )
	\left ( \Phi^{Ci} \Phi^{Dj} \Phi^{Mk} \Phi^{Nl} \right )
 	\left ( e^i_{\; \mu} e^j_{\; \nu} e^k_{\; \mu} e^l_{\; \nu} \right )
\label{3-2-60}
\end{eqnarray}
Unfortunatelly in this case it is impossible to simplify the quartic term in the consequence of the
specific form of the SU(3) structural constant $f^{ABC}$. 

\section{Summary}

In this paper we have applied the spin-charge separation for the SU(3) gauge field and have shown
that the SU(2) gauge field may have two different spin-charge separations. We have shown that ground
states in the SU(3) case can be divided into two
branches: the first one is similar to the SU(2) case, but the second branch contains special
vacuum states as there exist relations between eigenvalues of the matrix $A^B_\mu A^B_\mu$. The
existence of these vacuum states shows that the perturbative vacuum state of the SU(3) gauge theory
can be broken down to different vacuum states characterized by different gauge condensates
$A^B_\mu A^B_\mu$.

\section{Acknowledgments}

The author thanks D. Ebert, M. Mueller-Preussker and the other colleagues of the Particle Theory
Group of the Humboldt University for kind hospitality. This work has been supported by DAAD.

\end{document}